\documentstyle[aps,pra,amsmath,amssymb,psfig,floats]{revtex}
 
\newcommand\ave[1]{\langle#1\rangle}
\newcommand\bave[1]{\big\langle#1\big\rangle}

\newcommand\EC{E_{{}_{\rm{C}}}}
\newcommand\EJ{E_{{}_{\rm{J}}}}

\newcommand{\bp}{{\boldsymbol p}}

\begin{document}

 \wideabs{
\title{Quantum fluctuations in one-dimensional arrays of condensates}
\author{Alessandro Cuccoli, Andrea Fubini,
        Valerio Tognetti}
\address{Dipartimento di Fisica dell'Universit\`a di Firenze
    and Istituto Nazionale di Fisica della Materia (INFM),
    \\ Largo E. Fermi~2, I-50125 Firenze, Italy}
\author{Ruggero Vaia}
\address{Istituto di Elettronica Quantistica
    del Consiglio Nazionale delle Ricerche,
    via Panciatichi~56/30, I-50127 Firenze, Italy,
    \\ and Istituto Nazionale di Fisica della Materia (INFM)}
\date{August 8, 2001}
\maketitle

\begin{abstract}
The effects of quantum and thermal fluctuations upon the fringe
structure predicted to be observable in the momentum distribution of
coupled Bose-Einstein condensates are studied by the
effective-potential method. For a double-well trap, the coherence
factor recently introduced by Pitaevskii and Stringari
\protect{[}Phys. Rev. Lett. {\bf 87}, 180402 (2001)\protect{]} is calculated
using the effective potential approach and is found in good agreement
with their result. The calculations are extended to the case of a
one-dimensional array of condensates, showing that quantum effects are
essentially described through a simple renormalization of the energy
scale in the classical analytical expression for the fringe structure.
The consequences for the experimental observability are discussed.
\end{abstract}


} 


Recently, extensive theoretical work was devoted to the problem of the
coherence properties arising from the interaction among two or more
trapped
condensates~\cite{DalfovoGPS1999,PitaevskiiS1999,PitaevskiiS2001}.
These effects are experimentally accessible~\cite{experiments} and can
be considered of the same nature as in Josephson
junctions~\cite{Smerzi}. As a matter of fact, the condensates can be
characterized by their phases $\hat\varphi_i$ and atom numbers
$\hat{n}_i$, which are conjugate canonical variables satisfying
$[\hat{\varphi}_i,\hat{n}_j]=i\,\delta_{ij}$. The Hamiltonian looks
indeed like the Josephson one~\cite{Smerzi,Leggett2001}, namely
\begin{equation}
 \hat{\cal H} = \frac{\EC}4\sum_{i=1}^{N_{{}_{\rm{S}}}}
 \, \hat n_i^2
 -\EJ \sum_{\langle{ij}\rangle} \cos(\hat\varphi_{i}-\hat\varphi_j)\,,
\label{e.H}
\end{equation}
for an array of $N_{{}_{\rm{S}}}$ condensates, where the second
summation runs over the nearest-neighbor condensates. For two
interacting condensates ($i=1,2$) this reduces to
\begin{equation}
 \hat{\cal H}_0 = \frac{\EC}{2}~\hat n^2 - \EJ\cos\hat\varphi ~,
 \label{e.H0}
\end{equation}
where $\hat{n}=(\hat{n}_1-\hat{n}_2)/2$,
$\hat\varphi=\hat\varphi_1-\hat\varphi_2$, and the unessential
conserved term $(\EC/8)(\hat{n}_1{+}\hat{n}_2)^2$ is omitted. The
interaction parameters $\EC$ and $\EJ$ depend on the chemical potential
and the wave-function overlap of neighboring condensates,
respectively~\cite{Smerzi}. For the above Hamiltonians it is convenient
to define a quantum coupling parameter $g$ as the ratio between the
quantum energy scale $\hbar\omega_0=\sqrt{\EC\EJ}$ of the quasiharmonic
excitations and the overall energy scale of the nonlinear interaction,
$\EJ$, so that quantum effects are weak (strong) for
\begin{equation}
 g=\sqrt{\EC/\EJ}
\end{equation}
small (large) compared to $1$. The temperature is also conveniently
measured in the scale of $\EJ$ by the dimensionless parameter
$t\equiv{T}/\EJ$.

For the two-condensate case, starting from the uncertainty principle
$\ave{\Delta\hat{n}^2}~\ave{\Delta\sin^2\hat\varphi}\geq
(\hbar^2/4)~\ave{\cos\hat\varphi}$, the authors of
Ref.~\onlinecite{PitaevskiiS2001} introduced the {\em coherence factor}
as the thermodynamic average
\begin{equation}
 \alpha(g,t) = \ave{\cos \hat\varphi}~,
\label{e.alpha0}
\end{equation}
a parameter that can be directly related with the fringe structure in
momentum space, which is due to the onset of coherence between the
weakly linked condensates and can be detected by light scattering
experiments. Indeed, the observed momentum distribution $n(\bp)$ turns
out to be described, in terms of the single-condensate distribution
$n_0(\bp)$, as~\cite{PitaevskiiS1999,PitaevskiiS2001}
$n(\bp)=2\,[1+\alpha(g,t)\,\cos(p_xd/\hbar)]~n_0(\bp)$, where $d$ is
the distance of the condensate traps (assumed along the $x$~axis). In
an analogous way, purely quantum-mechanical correlations between
identical particles lead to a reduction of the neutron scattering
intensity of protons or deuterons which can be explained by a similar
mechanism~\cite{Lovesey2001}. At zero temperature, $1-\alpha$ measures
the decoherence effect of the only zero-point quantum fluctuations. The
value $\alpha=1$ occurs when the quantum coupling $g=0$ and corresponds
to the classical limit, while $\alpha$ decreases for increasing values
of $g$. When the temperature is finite, $\alpha$ decreases further due
to the thermal fluctuations that contribute to the destruction of
coherence. This thermal decoherence turned out to be significant even
at very low temperatures where the condensation starts to occur, as
shown in Ref.~\onlinecite{PitaevskiiS2001} for the single BEC junction.
However, the analogous calculation for an array of condensates, that
involves the phase correlation function
\begin{equation}
 G_r(g,t) = \bave{\cos(\hat\varphi_i-\hat\varphi_{i+r})}~,
\label{e.corr}
\end{equation}
cannot be performed by a direct numerical solution of the Schr\"odinger
equation, as done in Ref.~\onlinecite{PitaevskiiS2001}.

We note that an ideal tool to face such calculations is represented by
the effective potential method~\cite{GT1985-86}. Several
applications~\cite{CGTVV1995} proved indeed its usefulness: in
particular, it can be used for one- (1D) and two-dimensional (2D)
arrays~\cite{CFTV1999-2000}, where, otherwise, the exact quantum
solution can be obtained only by resorting to much heavier quantum
Monte Carlo simulations. The method is a semiclassical expansion in
terms of the quantum coupling $g=\sqrt{\EC/\EJ}$. The only pure-quantum
part of the fluctuations is considered in the self-consistent Gaussian
approximation so that the quantum Hamiltonian is reduced to a
classical-like one, where the parameters of the potential are
renormalized by the pure-quantum effects at any temperature. Of course
this renormalization vanishes at high temperatures,
($\beta=1/T\to{0}$), the approximation thus being more and more
accurate.

\medskip

\begin{figure}[tb]
 \centerline{\psfig{bbllx=16mm,bblly=10mm,bburx=182mm,bbury=135mm,%
figure=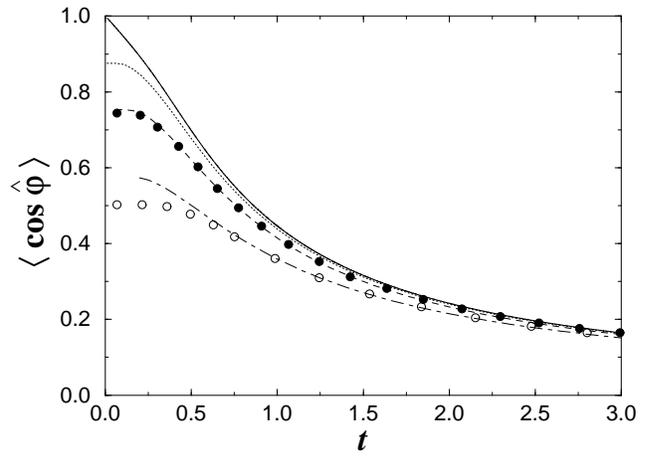,width=80mm,angle=0}}
 \caption{ Coherence factor $\alpha(g,t)=\ave{\cos\hat\varphi}$,
Eq.\eqref{e.alpha0}. Solid line, $g=0$ (classical); dotted line,
$g=0.5$; dashed line, $g=1$; dash-dotted line, $g=\sqrt{3}$. The
symbols are the exact result~\protect\cite{PitaevskiiS1999} for $g=1$
(full circles) and $g=\sqrt{3}$ (open circles). The latter value
corresponds to $\EC/\EJ=3$.}
\label{f.alpha0}
\end{figure}

Let us first consider the {\em two-condensate case}. The
renormalization parameter that takes into account the quantum
fluctuations only, is related to the pure-quan\-tum Gaussian spread of
the phase; it turns out to be
\begin{equation}
 D_0(g,t) = \Delta\hat\varphi^2_{\rm tot}-\Delta\hat\varphi_{\rm class}^2
 = \frac{g}{2\,\kappa}_0\Big(\coth f - \frac1{f}\,\Big)~,
\label{e.D0}
\end{equation}
where $f(g,t)=\beta\hbar\omega/2=g\kappa_0/(2\,t)$ and the renormalized
frequency is $\omega=\kappa_0\,\omega_0$; the {\em pure-quantum}
Hartree factor~\cite{CGTVV1995}
\begin{equation}
 \kappa_0(g,t) = e^{-D_0(g,t)/4}
\label{e.kappa0}
\end{equation}
is determined self-consistently with $D_0(g,t)$, which is a decreasing
function of temperature and vanishes for $t\to\infty$. The effective
classical Hamiltonian~\cite{CGTVV1995} corresponding to
Eq.~\eqref{e.H0} bears the same functional form,
\begin{equation}
 {\cal H}_0 = \frac{\EC}{2}~ n^2 - \EJ\, \kappa_0^2(g,t) \cos\varphi ~,
 \label{e.H0eff}
\end{equation}
with the renormalized Josephson coupling $\EJ~\kappa_0^2(g,t)$ (some
additive uniform terms, which do not affect the calculation of thermal
averages, have been neglected). In the effective potential formalism
any quantum average is evaluated by means of a classical-like
expression, that for the case of the coherence factor reads
\begin{eqnarray}
 \alpha(g,t) &=& \frac1{\cal Z_{{}_{\rm{C}}}} \int d\varphi~
 (\kappa_0^2 \cos\varphi)~\exp\Big[\frac{\kappa_0^2}t\,\cos\varphi\Big]~,
\notag\\
 &=& \kappa_0^2~{\cal{I}}_0({\kappa_0^2}/t) ~,
\label{e.alpha0eff}
\end{eqnarray}
where ${\cal{I}}_0$ is the logarithmic derivative of the modified
Bessel function of the first kind $I_0(x)$:
\begin{equation}
 {\cal I}_0(x)= \frac{d\ln I_0(x)}{dx}=\frac{I_1(x)}{I_0(x)} ~.
\label{e.I0}
\end{equation}
For $g=0$ one has $\kappa_0(0,t)=1$ and Eq.~\eqref{e.alpha0eff} reduces
to the exact classical coherence factor
$\alpha_{\rm{cl}}(t)={\cal{I}}_0(1/t)$, while it is worthwhile to
notice that within our approximation quantum effects are described by a
renormalization of the temperature scale. The results are reported for
different quantum-coupling values in Fig.~\ref{f.alpha0}, where
comparison with the exact ones is made. It appears that the method is
reliable as long as the renormalizations are small, $D_0\lesssim{1}$.
Note that $D_0$ is a decreasing function of $t$, so this condition is
verified also at strong coupling if $t$ is high enough, while the
validity in the whole temperature range requires $D_0(g,0)\lesssim{1}$,
i.e., $g\lesssim{1.5}$. The result for $g=1$ is in very good agreement
with the corresponding one of Ref.~\onlinecite{PitaevskiiS2001} in the
whole temperature range, while, as expected, the result for
$g=\sqrt{3}$ matches the exact data for $t\gtrsim{0.5}$. This testifies
to the reliability of the method, which can be then pursued also for
condensate arrays, where quantum fluctuations are expected to be
weaker.

\medskip

Let us now come to the case of a {\em 1D  condensate array}. The
effective Hamiltonian corresponding to Eq.~\eqref{e.H} reads
\begin{equation}
 {\cal H} = \sum_{i=1}^{N_{{}_{\rm{S}}}} \bigg[\frac{\EC}4\, n_i^2
 -\EJ\,\kappa^2(g,t) \cos(\varphi_{i}-\varphi_{i+1})\bigg]~,
\label{e.Heff}
\end{equation}
where now the pure-quantum Hartree factor is given by
\begin{equation}
 \kappa(g,t) = e^{-{\cal D}_1(g,t)/4} ~,
\label{e.kappa}
\end{equation}
and is to be self-consistently evaluated together with the
renormalization parameter
\begin{equation}
 {\cal D}_1(g,t) = \frac{g}{2\sqrt{2}\,\kappa}
 \frac1{N_{{}_{\rm{S}}}}\sum_k \Big|\sin\frac k2\,\Big|
 ~\Big(\coth f_k - \frac1{f_k}\,\Big) ~,
\label{e.D}
\end{equation}
which is a decreasing function of $t$ since
\begin{equation}
 f_k(g,t)=\frac{\beta\hbar\omega_k}2
 =\frac {g\,\kappa}{\sqrt{2}\,t}~\Big|\sin\frac k2\,\Big| ~;
\end{equation}
indeed, the renormalized dispersion relation takes the form
$\hbar\omega_k=\sqrt{2}\,\kappa\,\omega_0\,\big|\sin(k/2)\big|$. The
approximation holds if ${\cal D}_1(g,t)\lesssim{1}$, and since at zero
temperature ${\cal{D}}_1(g,0)=g/(\sqrt{2}\pi\kappa)$, the condition
amounts to require $g\lesssim{3.5}$.

For the correlation function \eqref{e.corr} the effective-potential
recipe gives~\cite{CTV1991}
\begin{equation}
 G_r(g,t) = e^{-{\cal D}_r/2}~
 \big[\,{\cal I}_0(\kappa^2/t)\big]^{|r|} ~,
\label{e.Greff}
\end{equation}
with ${\cal I}_0(x)$ as in Eq.~\eqref{e.I0}; the analytical evaluation
uses periodic boundary conditions and the translation invariance of the
potential~\cite{CTV1991}. The renormalization parameters for the $r$th
neighbors read
\begin{equation}
 {\cal D}_r(g,t)
 = \frac{g}{2\sqrt{2}\,\kappa}\frac1{N_{{}_{\rm{S}}}}\sum_k
 \frac{\sin^2 (rk/2)}{\big|\sin (k/2)\big|}
 \,\Big(\coth f_k {-} \frac1{f_k}\,\Big) \,;
\label{e.Dr}
\end{equation}
for $r\to\infty$ one can replace $\sin^2(rk/2)$ by its average $1/2$
and get the asymptotic value ${\cal{D}}_\infty(g,t)$, which is
logarithmically divergent for $t=0$. In particular, one has the
power-law asymptotic behavior
$e^{-{\cal{D}}_\infty(g,t)/2}\sim{t^\eta}$, with
$\eta=g/[4\sqrt{2}\pi\kappa(g,0)]={\cal{D}}_1(g,0)/4$. Moreover, the
zero-$t$ correlation function behaves as
$G_r(g,0)=e^{-{\cal{D}}_r(g,0)/2}\sim{r^{-\eta}}$ in agreement with the
theoretical prediction for the low-coupling phase~\cite{BradleyD1984}.

\begin{figure}[tb]
 \centerline{\psfig{bbllx=16mm,bblly=10mm,bburx=182mm,bbury=135mm,%
figure=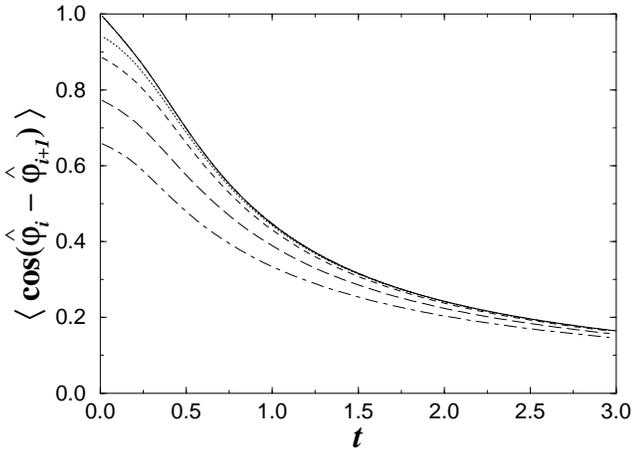,width=80mm,angle=0}}
 \caption{Nearest-neighbor correlation function of the 1D condensate
chain $G_1(g,t)=\bave{\cos(\hat\varphi_i-\hat\varphi_{i+1})}$,
Eq.\eqref{e.alphaeff}. Solid line, $g=0$ (classical); dotted line,
$g=0.5$; short-dashed line, $g=1$ ; long-dashed line, $g=2$;
dash-dotted line, $g=3$.}
 \label{f.alpha}
\end{figure}

For the first-neighbor correlation, we can define the analog of the
coherence factor, Eqs.~\eqref{e.alpha0}, which in the
effective-potential approximation bears the same form of
Eq.~\eqref{e.alpha0eff},
\begin{equation}
 \bave{\cos(\hat\varphi_i-\hat\varphi_{i+1})}
 = \kappa^2~{\cal I}_0\big(\kappa^2/t\big)~;
\label{e.alphaeff}
\end{equation}
it is plotted in Fig.~\ref{f.alpha}. Comparing with the two-condensate
case (e.g., curves for $g=0.5$ and $1$), it appears that the quantum
correction is weaker. It is worthwhile to notice that in the classical
case, i.e., $g=0$, Eq.~\eqref{e.Greff} reduces to the correct classical
expression.

The formulas above can also be easily generalized~\cite{CFTV1999-2000}
to the 2D case, of recent experimental
realization~\cite{GreinerBMHE2001}, but the involved classical-like
integrals cannot be evaluated analytically as in Eq.~\eqref{e.Greff},
so a numerical approach is needed, e.g., Monte Carlo (MC) simulation
(still much easier than quantum MC).

For the multicondensate array (with spacing $d$ along the
$x$~direction), the momentum distribution
$n({\bp})=N_{{}_{\rm{S}}}\,n_0({\bp})~\Xi(p_x)$ exhibits a fringe
structure determined by the structure function
\begin{eqnarray}
 \Xi(p_x) &\equiv&  {\sum}_r~ e^{-irp_x d/\hbar}
  ~\bave{\cos(\hat{\varphi}_i-\hat{\varphi}_{i+r})}
\notag\\
 &=&  {\sum}_r~ e^{-irp_x d/\hbar} \, e^{-{\cal D}_r/2} \,
 \big[{\cal I}_0(\kappa^2/t)\big]^{|r|}~,
\end{eqnarray}
where Eq.~\eqref{e.Greff} has been inserted; clearly,
$\Xi(p_x)=\Xi(p_x+2\pi\hbar/d)$ is periodic. In this expression one can
take advantage of the fact that the renormalization coefficients
${\cal{D}}_r(g,t)$ rapidly converge to the asymptotic value
${\cal{D}}_\infty(g,t)$, separating it as
\begin{eqnarray}
 \Xi(p_x) &=& \Xi_0(p_x) + \Xi_1(p_x)
\label{e.Xi}
\\
 \Xi_0(p_x) &=& e^{-{\cal D}_\infty/2}~
 \frac{1-\alpha^2}{1+\alpha^2- 2\alpha\cos(p_x d /\hbar)}~,
\label{e.Xi0}
\\
 \Xi_1(p_x) &=& {\sum}_r~ e^{-ir\,p_x d/\hbar}\,
 \alpha^{|r|} \big(e^{-{\cal D}_r/2}-e^{-{\cal D}_\infty/2}\big)~,
\label{e.Xi1}
\end{eqnarray}
where the quantity
\begin{equation}
 \alpha(g,t) = {\cal I}_0\big({\kappa^2(g,t)}/t\big)~,
\end{equation}
characterizes the coherence effect. Looking at Eq.~\eqref{e.Xi1} it
appears that the difference in parentheses cuts off the contribution
from large values of $r$, so this term does not show sharp features and
the fringe structure is mostly given by Eq.~\eqref{e.Xi0}. In the
classical limit ($g=0$) ${\cal{D}}_r=0$ vanishes and so does
$\Xi_1(p_x)$, and the exact classical result~\cite{PitaevskiiS2001} is
immediately recovered since $\alpha(0,t)=\alpha_{\rm{cl}}(t)$.

\begin{figure}[tb]
 \centerline{\psfig{bbllx=16mm,bblly=10mm,bburx=182mm,bbury=135mm,%
figure=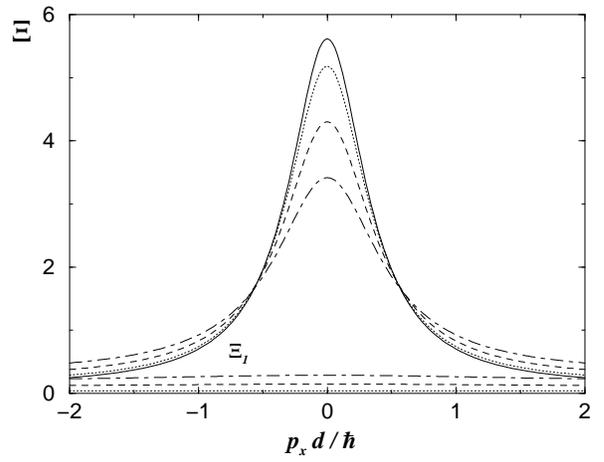,width=80mm,angle=0}}
 \caption{ Structure function $\Xi(p_x;g,t)$ of the 1D condensate chain
[Eq.\eqref{e.Xi}] at fixed $t=0.5$.  Solid line, $g=0$ (classical);
dotted line, $g=1$; dashed line, $g=2$; dash-dotted line, $g=3$. The
low-lying curves report the small contribution of the term
$\Xi_1(p_x;g,t)$ for the same values of $g$. One single periodic
interval is displayed.}
 \label{f.xig}
\end{figure}

\begin{figure}[tb]
 \centerline{\psfig{bbllx=16mm,bblly=10mm,bburx=182mm,bbury=135mm,%
figure=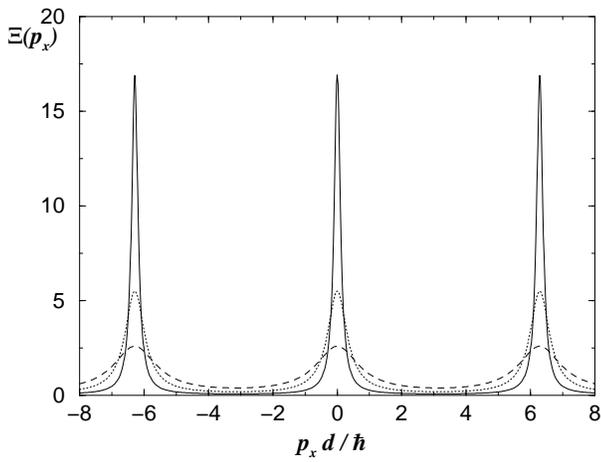,width=80mm,angle=0}}
 \caption{ Structure function $\Xi(p_x;g,t)$ of the 1D condensate chain
[Eq.\eqref{e.Xi}], at fixed $g=0.5$. Solid line, $t=0.2$; dotted line,
$t=0.5$; dashed line, $t=1$.}
 \label{f.xit}
\end{figure}

Representative results for the full structure function $\Xi(p_x)$ and
the correction $\Xi_1(p_x)$ are reported in Fig.~\ref{f.xig} for some
values of the quantum coupling $g$ at fixed temperature $t=0.5$. The
stronger the quantum effects (i.e., high $g$ and low $t$) the higher is
$\Xi_1(p_x)$, but in the considered parameter range the relative
contribution of $\Xi_1(p_x)$ is small and broad, confirming that the
main structure is contained in $\Xi_0(p_x)$. In Fig.~\ref{f.xit} the
structure function $\Xi(p_x)$ is plotted for fixed $g=0.5$ and
different temperatures, in a wider range of $p_x$ so to display the
periodic behavior.

\begin{figure}[tb]
 \centerline{\psfig{bbllx=17mm,bblly=70mm,bburx=200mm,bbury=212mm,%
figure=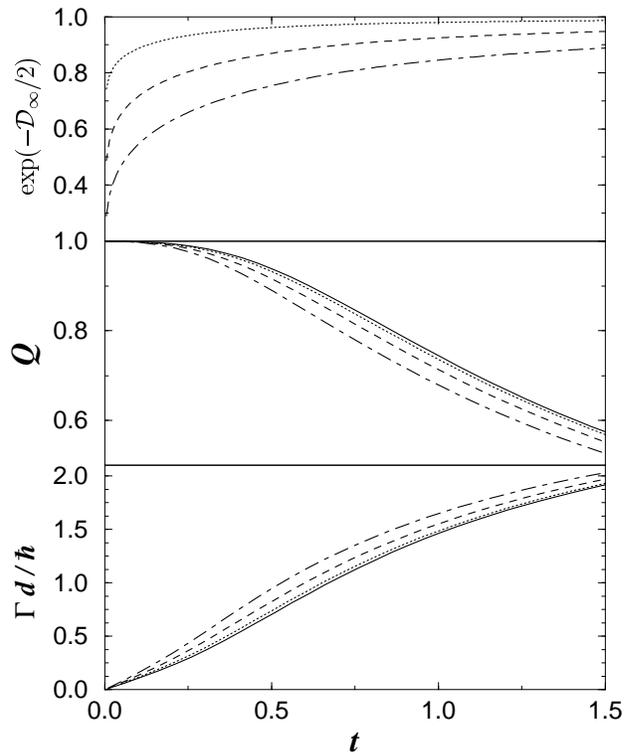,width=80mm,angle=270}}
 \caption{Characteristics of the fringe structure $\Xi(p_x;g,t)$ of the
1D condensate array vs temperature and for different values of the
quantum coupling, $g=0$ (solid lines, classical), $g=1$ (dotted lines),
$g=2$ (dashed lines), and $g=3$ (dash-dotted lines). Top: quantum
attenuation factor $e^{-{\cal{D}}_\infty/2}$. Middle: contrast factor
$Q(g,t)$, Eq.~\eqref{e.Q}. Bottom: peak width $\Gamma(g,t)$,
Eq.~\eqref{e.Gamma}.}
 \label{f.QGamma}
\end{figure}

Finally, let us discuss how phase fluctuations do affect the observable
fringe structure. Neglecting the contribution of $\Xi_1(p_x)$, the
structure function shows maxima
$\Xi_{\rm{M}}=e^{-{\cal{D}}_\infty/2}\,(1+\alpha)/(1-\alpha)$ around
the points $p_x=2\ell\pi\hbar/d$ ($\ell$ integer), and minima
$\Xi_{\rm{m}}=e^{-{\cal{D}}_\infty/2}\,(1-\alpha)/(1+\alpha)$ in
between. The contrast factor is defined as
\begin{equation}
 Q(g,t) = (\Xi_{\rm{M}}-\Xi_{\rm{m}})/(\Xi_{\rm{M}}+\Xi_{\rm{m}})
\label{e.Q}
\end{equation}
($Q\sim{1}$ means good contrast), while the peak width $\Gamma$,
defined as the full width at half height between maxima and minima, is
\begin{equation}
 \Gamma(g,t) = (4\hbar/d)~
 \tan^{-1}\big[(1-\alpha)/(1+\alpha)\big] ~.
\label{e.Gamma}
\end{equation}
For $\Gamma\ll{2\pi\hbar/d}$, i.e., when $\alpha$ is not far from its
zero-temperature value $\alpha(g,0)=1$, the peaks are sharp and have a
quasi-Lorentzian shape. However, Eq.~\eqref{e.Xi0} shows that the
overall intensity is depressed by the factor $e^{-{\cal{D}}_\infty/2}$,
which is plotted in Fig.~\ref{f.QGamma} together with $Q$ and $\Gamma$,
for different coupling values: it appears that the main effect of
quantum fluctuations is to weaken the overall intensity, while the
contrast and the peak width are predominantly driven by temperature.

\medskip

The authors wish to thank L.~Pitaevskii and S.~Strin\-gari, as well as
F.~S. Cataliotti, F.~Ferlaino, C.~Fort, and M.~Inguscio for useful
discussions. This research is supported by the COFIN2000-MURST fund.

\end{document}